\documentclass[a4paper,10pt]{article}
\usepackage[utf8]{inputenc}
\usepackage{amsmath, amsthm, amssymb}
\usepackage{hyperref}
\usepackage{bm}
\usepackage{relsize}
\usepackage{authblk}
\usepackage{dcolumn}
\usepackage{setspace}
\usepackage[normalem]{ulem}
\usepackage{ifpdf}
\usepackage{physics}
\usepackage{cite}

\title{Triviality of Entanglement Entropy in the Galilean Vacuum}
\author{Itamar Hason\thanks{itamarhason@gmail.com}}
\affil{Raymond and Beverly Sackler School of Physics and Astronomy, \\ Tel Aviv University, Tel Aviv 69978, Israel}
\date{\today}

\begin{document}

\maketitle

\begin{abstract}
We study the entanglement entropy of the vacuum in non-relativistic local theories with Galilean or Schr\"{o}dinger symmetry. We clear some confusion in the literature on the free Schr\"odinger case. We find that with only positive $U(1)$ charge particles (states) and a unique zero $U(1)$ charge state (the vacuum) the entanglement entropy must vanish in that state.
\end{abstract}

\section{Introduction}

Entanglement entropy is a property of quantum systems described by a Hilbert space. Given a state $\left| \psi \right>$, one defines the entanglement entropy between two subsystems $A$ and $B$ to be the von-Neumann entropy of the density matrix $\rho = \left|\psi\right> \left<\psi\right| $ traced over one of the subsystems. Entanglement entropy has been intensively studied in relativistic theories in which the area law has been demonstrated explicitly \cite{Srednicki:1993im} and the holographic interpretation of entanglement was founded \cite{Ryu:2006bv}.

Galilean field theories are theories where the spacetime symmetries are Galilean rather than Lorentzian. Recall that the Galilean algebra contains a central charge $M$ generating the particle number symmetry. It is related to the other spacetime symmetries by the commutator
\begin{equation}
 \left[ P_i, K_j \right] = -i \delta_{ij} M \ .
\end{equation}
This central charge is responsible for many Galilean phenomena different from what we are used to in relativistic field theories \cite{Jensen:2014aia}. See \cite{Nishida:2007pj,Nishida:2010tm,Balasubramanian:2008dm,Goldberger:2014hca,Arav:2014goa, Arav:2016xjc, Arav:2016akx, Auzzi:2016lrq, Pal:2016rpz, Baggio:2011ha, Barvinsky:2017mal} for related recent Galilean and Schr\"odinger field theory papers.

Recently, the interest in entanglement entropy in Galilean field theories has raised and a few works have been published in which the entanglement entropy is computed in Galilean framework using different methods. In particular, in \cite{Solodukhin:2009sk} a computation using the heat-kernel method and an argument using a Lifshitz holographic dual are given for the case of free Schr\"{o}dinger field theory. We suspect both arguments are not appropriate for the free Schr\"odinger operator. The first due to an ill-defined Schr\"{o}dinger operator and the second by using a non-Schr\"{o}dinger dual. Note that a later work \cite{Auzzi:2017wwc} refer \cite{Solodukhin:2009sk} going forward with more free Schr\"odinger computations, based on the same inappropriate method. It's possible that the computations done in \cite{Solodukhin:2009sk} could be used to study a different case, but not the free Schr\"odinger case. The problem with the operator in \cite{Solodukhin:2009sk} was mentioned briefly in \cite{Pal:2017ntk} and some of the fundamental ingredients leading to the correct result for the free case were mentioned there as well, and in this paper we would like to elaborate on that. We present arguments for the triviality of the entanglement entropy under certain conditions in the Galilean vacuum and emphasize the importance of the particle number symmetry generator $M$.

\section{Free Schr\"{o}dinger}
We claim that the entanglement entropy of a subset of space in the Schr\"{o}dinger vacuum state is zero. Recall that given a representation of the Hilbert space as a product of two Hilbert spaces
\begin{equation}
 \mathcal{H} = \mathcal{H}_A \otimes \mathcal{H}_B
\end{equation}
and given a (pure) state $\left| \psi \right>$ in $\mathcal{H}$, the entanglement entropy of $\left| \psi \right>$ with respect to $A$ (or $B$) is defined to be 
\begin{equation}
 S(A) = -\text{Tr} \left( \rho_A \text{log} \rho_A \right)
\end{equation}
where $\rho_A$ is the density matrix $\rho = \left| \psi \right> \left< \psi \right|$ reduced to the subspace $\mathcal{H}_A$ by tracing over the complement subspace $\mathcal{H}_B$. Note that the entanglement entropy is defined on a fixed time.

First, we claim that if the two subspaces $\mathcal{H}_A$ and $\mathcal{H}_B$ are completely uncorrelated on the given state $\left| \psi \right>$, then the entanglement entropy should vanish. By complete uncorrelation we mean that every correlation function that involves operators defined on either $\mathcal{H}_A$ or $\mathcal{H}_B$ is given by the product of the correlation functions on $\mathcal{H}_A$ and $\mathcal{H}_B$ separately.

Second, in the free Schr\"{o}dinger theory, the equal time two point function vanishes on separated points, indeed, in the free case it is well known that
\begin{equation}
 \left< \phi(\vec{x}_1, t) \phi(\vec{x}_2, t) \right> \sim \delta(\vec{x}_1-\vec{x}_2) \ .
\end{equation}

Third, we claim that every $n$-point function can be factorized to the $A$ part and the $B$ part. We can use Wick's theorem to write the $n$-point function as a sum of products of two point functions. Every term that involves a two point function that mixes $A$ and $B$ necessarily vanishes because $A\cap B = \emptyset$. Therefore, any $n$-point function may be factorized to the $n_A$- and $n_B$-point function, i.e., $n$-point functions are separable.

Therefore, we conclude that the free Schr\"{o}dinger vacuum has zero entanglement entropy for every subset $A$.

Actually \cite{Pal:2017ntk}, the fundamental reason for the vacuum state in the free Schr\"{o}dinger field theory to be entanglement free is that the Hilbert space has a basis in terms of a set of particles localized in space
\begin{equation}
 \left| \vec{x}_1, \vec{x}_2, ..., \vec{x}_n \right> \ .
\end{equation}
The vacuum state is the state with no particles, or with zero $U(1)$ charge, and it is the only such state. In comparison, relativistic field theory doesn't have states with completely localized particles.

We can look at a subspace $A$ and there we also have such a basis provided that $\vec{x}_i \in A$, and similarly for $B$. Therefore, the vacuum state of the full space can be written as $\left| 0 \right> = \left| 0 \right>_A \otimes \left| 0 \right>_B$, and thus, obviously, the vacuum state is not entangled, tracing over $B$ leaves us with a pure state $\left| 0 \right>_A$.

\section{Theories with Galilean Symmetry}
From the above argument we should expect that a Schr\"{o}dinger theory is entanglement free in the vacuum state if the vacuum has, and is the only state to have, zero $U(1)$ charge.

We want to generalize the above correlation functions argument to not-necessarily free theories.
Note that the two point function must satisfy
\begin{equation}
 \left< \phi(\vec{x}_1, t_1) \phi(\vec{x}_2, t_2) \right> = e^{\frac{im}{2}\frac{\left(\vec{x}_2-\vec{x}_1\right)^2}{t_2-t_1}} f(t_2-t_1)
\end{equation}
To prove that, we shall use space and time translation invariance as well as boost invariance (of the theory and of the vacuum) \cite{Golkar:2014mwa}.

This expression is not well defined when $t_2=t_1$. To see that this is zero for $t_2=t_1$ on separated points $\vec{x}_1\neq \vec{x}_2$ we can regularize the space dependence by integrating over a small region of $\vec{x}_2-\vec{x}_1$ and take the limit $t_2-t_1 \rightarrow 0$ \footnote{It's important that we first take the limit $t_2-t_1 \rightarrow 0$ and only then we take the regularization parameter, i.e., the space integration region, to zero.}. When we do that, unless $\vec{x}_2=\vec{x}_1$ and provided that the function $f$ diverges polynomially, we get zero -- since the exponent phase varies rapidly, if one integrates this phase factor around a small region of $x$s (thus, inserting a regulator to find the limit of equal times) multiplied by a (diverging) power function, the integral would vanish (in the limit of equal times).

Generalizing to $n$-point functions isn't trivial because in the non-free case we have loop integrals over the whole spacetime. Indeed, we must use some knowledge about the particle density in the problem, otherwise, using non-zero chemical potential, one may construct examples with non-zero entanglement entropy.

We'll therefore adopt a more general approach. Let's prove that in a local field theory with Galilean symmetry, with only positive $U(1)$ charge particles (states), and a unique state with zero $U(1)$ charge (the vacuum), there is no entanglement entropy in that state.

If the decomposition of the vacuum to a superposition of product states on $A$ and $B$ is $\left| 0 \right> = \left| 0 \right>_A \left| 0 \right>_B$ then clearly there is no entanglement, because after tracing over $B$ we get the pure state $\left| 0 \right>_A$. For that not to be the case one must have a non-trivial decomposition $\left| 0 \right> = \sum { \left| i \right>_A \left| j \right>_B }$ where $\left| i \right>_A$ and $\left| j \right>_B$ some states in the Hilbert spaces $\mathcal{H}_A$ and $\mathcal{H}_B$ respectively. By charge decomposition \footnote{Here we use charge decomposition which is a direct result of a charge density existence assumption which is common in Galilean field theories. For example, in the free case, $M_A$ and $M_B$ can be easily defined algebraically via their eigenvalues in the basis of localized particles -- a localized particles state with $n_A$ particles in region $A$ and $n_B$ particles in region $B$ is an eigenstate of $M_A$ and $M_B$ with eigenvalues $n_A$ and $n_B$ respectively. $M_\text{boundary}$ is mentioned in order to allow for boundary charges, it should be understood that it vanishes in case nothing like that exists (or in case one chooses to include the boundary in $A$ or $B$).} one has
\begin{equation}
M_0 = M_i + M_j + M_\text{boundary}
\end{equation}
but since $M_0 = 0$ and $M$ is non-negative (it must be non-negative on $A$, $B$ and the boundary as well), one must have $M_i = M_j = 0$, and since there is a unique\footnote{It has to be unique on $A$ and $B$ as well, otherwise we could have built multiple products $\left| 0 \right>_A \left| 0 \right>_B$ in contradiction to the uniqueness of $\left| 0 \right>$.} $U(1)$ charge state, one gets $\left| 0 \right> = \left| 0 \right>_A \left| 0 \right>_B$.

Maybe as an explanatory example\footnote{The free Schr\"odinger case is the simplest case but it is not the only case obeying the above assumptions. Indeed, adding interactions to this free theory doesn't change the simple positivity condition or the uniqueness of a state free of particles and therefore with zero charge. Further examples include a free field with external potential (with positive spectrum), i.e. with the added term $\psi^\dagger \psi V$, or the nonlinear Schr\"odinger field theory with the interaction term $\lambda \psi^\dagger \psi^\dagger \psi \psi$, both of which keep the charge properties of all states invariant.}, we can look again at the free case. When we decompose the Hilbert space to $\mathcal{H}_A$ and $\mathcal{H}_B$, the basis for these spaces is formed of localized particles in $A$ and localized particles in $B$ (the boundary may be taken separately but we will avoid this unnecessary complication here). The $M$ charge for every state can be written as the sum of $M_A$ and $M_B$ (and $M_\text{boundary}$) all of which must be non-negative. There is also uniqueness of $M=0$ states in $A$, $B$ (and the boundary), so the vacuum must be just the product of the two vacua which proves entanglement freedom of the vacuum.

\section*{Acknowledgments}
We would like to thank Igal Arav and Yaron Oz for valuable discussions and the reviewers for significant comments. This work is supported in part by the I-CORE program of Planning and Budgeting Committee (grant number 1937/12), the US-Israel Binational Science Foundation, GIF and the ISF Center of Excellence.

\end{document}